\documentclass[aps,prl,10pt,twocolumn,superscriptaddress,showpacs]{revtex4-1}
\usepackage{graphicx}
\usepackage{dcolumn}
\usepackage{bm}

\begin{document}

\title{Evaluation of the curvature-correction term from the equation of state of nuclear matter}
\author{K. V. Cherevko}
\email{konstantin.cherevko@gmail.com}
  \affiliation{College of Nuclear Science and Technology, Beijing Normal University, Beijing 100875, China}
  \affiliation{Beijing Radiation Center, Beijing 100875, China}
  \affiliation{Physics Faculty, Taras Shevchenko National University of Kyiv, 4 Glushkova av., Kyiv, 03022, Ukraine}
\author{L. A. Bulavin}
  \affiliation{Physics Faculty, Taras Shevchenko National University of Kyiv, 4 Glushkova av., Kyiv, 03022, Ukraine}
\author{L. L. Jenkovszky}
\email{jenk@bitp.kiev.ua}
  \affiliation{Bogolyubov Institute for Theoretical Physics (BITP), Ukrainian National Academy of Sciences, Metrolohichna str. 14-b, Kyiv, 03680, Ukraine}
\author{V. M. Sysoev}
  \affiliation{Physics Faculty, Taras Shevchenko National University of Kyiv, 4 Glushkova av., Kyiv, 03022, Ukraine}
\author{F.-S. Zhang}
 \email{fszhang@bnu.edu.cn}
  \affiliation{College of Nuclear Science and Technology, Beijing Normal University, Beijing 100875, China}
  \affiliation{Beijing Radiation Center, Beijing 100875, China}
  \affiliation{Center of Theoretical Nuclear Physics, National Laboratory of Heavy Ion Accelerator of Lanzhou, Lanzhou 730000, China}

\date{\today}

\begin{abstract}
Based on the nuclear equation of state, the curvature correction term to the surface tension coefficient is calculated. Tolman's $\delta$ correction is shown to be sensitive to the Skyrme force parametrization. The temperature dependence of the Tolman length, important in heavy ion collisions experiments, is derived. In suggested approach the curvature term is related to the bulk properties of the nuclear matter through the equation of state. The obtained results are compared with the existing theoretical calculations based on the Gibbs-Tolman formalism and with the theoretical predictions concerning its dependence on the interparticle distance.
\end{abstract}\texttt{}


\maketitle

\textit{Introduction.---}During recent decades the rapid progress has been made in the development of the macroscopic description of the nuclear matter \cite{Brack1985, Chomaz2005, *Moretto2011}. A number of works devoted to the thermodynamics of small systems or hydrodynamics of nuclear matter appeared \cite{ Baumgardt1975, *Siemens1979,*Wong2008, Gross2001, *Cherevko2011, Chomaz2004}. Among the macroscopic models used in nuclear physics a special role belongs to  theories based on the Droplet model of nuclei \cite{Myers1969}. They make possible the description of the average properties of a saturated system, such as a nucleus, consisting of two components (neutrons and protons), with account for the boundary effects and the presence of the diffuse layer. For nuclei with the mass number $A$ in studies of their surface properties the accounting for the curvature effects is important, which means the inclusion of additional terms of order $A^{\frac{1}{3}}$ in any expansion concerning the nuclear properties in terms of the fundamental dimensionless ratio between the interparticle spacing the nuclear radius, given by $A^{-\frac{1}{3}}$.

Corrections due to curvature may play an important role when studying light nuclei or processes where surface terms are important. Particularly important are they in the interpretation of multifragmentation experiments \cite{Gross1990, *Zhang1996, *Zhang1998, Chomaz2004}, where light nuclei necessarily appear, and the exponential dependence of the yield of fragments on the surface tension makes the process sensitive to the curvature corrections \cite{Elliott2013,*Toke2003}. The other important phenomena that may be affected by the changes introduced to the surface tension by the curvature corrections are the appearance of the neck region in the fission processes and hydrodynamic instability of the structures formed in heavy ion experiments, governed by surface effects \cite{Brosa1983, *Cherevko2014}. Quite a number of works exist devoted to studies of the surface energy and the  properties of the surfaces in nuclear matter \cite{Ravenhall1983, Boyko1990, Jenkovszky1994}. Furthermore, the definition of the dependence of the surface tension (and surface energy) on the surface curvature and studies of its influence on different physical properties is also being studied by various groups of authors \cite{Moretto2012, Pomorski2003}. Still, for decades it remains one of the most controversial issues in mesoscopic thermodynamics \cite{Anisimov2007, Blokhius2006, Kolomietz2012}.

To set the stage, we recall that the thermodynamic description of the curvature correction, originating from the difference between the equimolar surface and the surface of tension \cite{Rowlinson1982, Rowlinson1994}, dates back to 1940s. The Tolman length $\delta$ was originally introduced in Ref. \cite{Tolman1949} to describe the curvature dependence of the surface tension of a small liquid droplet. It was defined as a correction term in the surface tension $\sigma$ of the liquid-vapour droplet in the isothermal case:
\begin{equation}
\label{eq1}
   \sigma(R)=\sigma_\infty\left(1-\frac{2\delta}{R}+\cdot\cdot\cdot\right),
 \end{equation}
where $R$ is the droplet radius, equal to the radius of the surface tension \cite{Rowlinson1982, Rowlinson1994}, and $\sigma_\infty$ is the surface tension of the planar interface. Eq. (\ref{eq1}) originates from the Gibbs-Tolman-Kenig-Buff's thermodynamic equation and the  assumption that $\delta$ is independent of $R$ for $\delta\ll{R}$ \cite{Ono1960}. This physics should work not only for liquid droplets but also for any system with curved interface of a non negligible boundary layer \cite{Anisimov2007}. This situation corresponds to nuclei and nuclear systems with a finite diffuse layer \cite{Brack1985}. The value of Tolman length has the same order of magnitude as the average interparticle distance $r_0$ \cite{Gorski1989, Fisher1984}, that for a nuclear systems is $r_0\sim0.7$ fm at normal density $\rho\sim0.17$ fm$^{-3}$. Hence, mathematically the term $\frac{2\delta}{R}$ in Eq. (\ref{eq1}) becomes important for the systems with $R<14$ fm that means it is definitely important for nuclear systems with $R<8$ fm. The above estimations show that this approximation works well in a wide range of radii, and can play an important role for all known nuclei and structures that are formed in the heavy ion collisions experiments.

Within the above approximation, the dependence of the surface tension on the curvature of the interface is defined only by the Tolman length $\delta$. Therefore, the knowledge (evaluation) of $\delta$ is quite important. However, the sign of the known (calculated) values of the Tolman length are not unique: both negative and positive values can be found in the literature \cite{Blokhius2006, Rowlinson1982, Ono1960}. At the same time, there are no reliable experimental methods to evaluate it. The aim of this Letter is to introduce a method allowing the  evaluation of $\delta$ from the experimental data \cite{Bulavin1998}.

In studying the curvature-correction term for the nuclear matter one should keep in mind the connection between the surface and bulk properties \cite{Myers1969, Moretto2012}. As shown in the Droplet model, the coefficients in the term proportional to $A^{\frac{1}{3}}$ in the expansion of nuclear properties in terms of the fundamental dimensionless ratio $A^{-\frac{1}{3}}$ are connected to the bulk properties of the nuclear matter, described by terms proportional to $A$ and $A^{\frac{2}{3}}$. It justifies the approach suggested in this Letter to evaluate the curvature correction (Tolman's length $\delta$) from the equation of state (EOS) of nuclear matter.

\begin{figure}[h]
\includegraphics[scale=1.0]{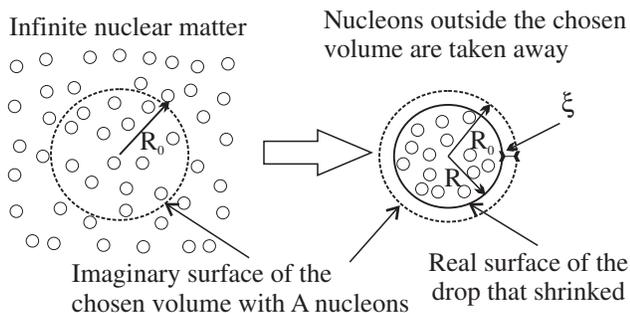}
\caption{Schematic picture of the {\it gedanken} experiment} \label{fig1}
\end{figure}

\textit{Theoretical model.---}
Let us consider infinite nuclear matter ($P_0$, $T=const$) with the chosen spherical volume $V_0=\frac{4}{3}{\pi}{R_0}^3$ in it, consisting of $A$ nucleons. Next, one may perform the following {\it gedanken} experiment. If all the nucleons outside the chosen volume are removed, one gets a {\textquotedblleft}nuclear droplet{\textquotedblright} that, due to surface tension, shrinks to the volume $V=\frac{4}{3}{\pi}{R}^3$, where $R$ is the final radius of the chosen volume [Fig. \ref{fig1}].
This droplet remains in equilibrium in one of the following regimes: either when the timescale of the particles evaporation is big enough and the evaporated particles are removed from the surface ($P(V,T)=0$), or when the {\textquotedblleft}nuclear liquid{\textquotedblright} is surrounded by the saturated {\textquotedblleft}nuclear vapour{\textquotedblright} with $P^{liq}=P^{vap}$ and chemical potential $\mu^{liq}=\mu^{vap}$. Th seuperscripts {\textit{liq}} and {\textit{vap}} denote liquid and vapor phases respectively.

Another {\it gedanken} experiment is to consider a big nucleus ${V_0}^{big}=2V_0=2\left(\frac{4}{3}{\pi}{R_0}^3\right)$ in equilibrium, consisting of $2A$ nucleons, subsequently divided in two equal parts $V$ with $A$ nucleons each. Due to the surface effects, the smaller nuclei will shrink, so that $2V\left(\frac{4}{3}{\pi}{R}^3\right)<{V_0}^{big}$.

Next we introduce the parameter $\xi=R_0-R>0$ [Fig. \ref{fig1}] assumed to be independent of $R$. Let us consider the equation of state (EOS) to be of the \cite{Bulavin1998}:
\begin{equation}
\label{eq2}
   \Delta{P}=P-P_0=f\left(\frac{V_0-V}{V_0}\right),
\end{equation}
where $P_0$, $V_0$ and $P$, $V$ are the initial and final points of the system evolution along the coexistence curve. Within our {\it gendanken} experiment, a small parameter $\frac{\xi}{R}$ can be introduced and, therefore the function $f$ from Eq. (\ref{eq2}) with the argument $\frac{V_0-V}{V_0}=1-\left[1+\frac{\xi}{R}\right]^{-3}$ can be expanded into the following series:
\begin{equation}
\label{eq3}
 \begin{array}{l}
   \Delta{P}=f(0)+\frac{\partial{f(0)}}{\partial\left(\frac{1}{R}\right)}\frac{1}{R}+\frac{1}{2}\frac{\partial^2{f(0)}}{\partial\left(\frac{1}{R}\right)^2}\left(\frac{1}{R}\right)^2+\cdot\cdot\cdot\\
   =f(0)+3\dot{f}(0)\left(\frac{\xi}{R}\right)+\frac{1}{2}\left(9\ddot{f}(0)-12\dot{f}(0)\right)\left(\frac{\xi}{R}\right)^2+\cdot\cdot\cdot.
 \end{array}
\end{equation} On the other hand, the excess pressure due to the surface tension in the left hand side of the Eq. (\ref{eq3}) can be found from the Laplace equation \cite{Gibbs1928}:
\begin{equation}
\label{eq4}
   \Delta{P}=\frac{2\sigma(R)}{R},
\end{equation}
By substituting $\sigma(R)$ from Eq. (\ref{eq1}) to Eq. (\ref{eq4}), and limiting ourself by the second order in the expansion (\ref{eq3}), and equating the coefficients of the similar order $\frac{1}{R}$, one gets from Eqs. (\ref{eq3}) and (\ref{eq4})
\begin{equation}
\label{eq5}
   \delta=\left[\frac{4\dot{f}(0)-3\ddot{f}(0)}{6\left(\dot{f}(0)\right)^2}\right]\sigma_\infty.
\end{equation}
To get numerical results for the Tolman length from Eq. (\ref{eq5}), it is necessary to chose the appropriate EOS of nuclear matter. In the present work we adopt the EOS of nuclear matter of low-temperature and high-density limit where $\lambda^3\rho\gg1$, that is, when the average de Broglie thermal wavelength $\lambda$ is larger than the average interparticle separation $\rho^{-\frac{1}{3}}$ in a form \cite{Lee2008}:
\begin{equation}
\label{eq6}
  \begin{array}{l}
   P(\rho_q,T)=\sum\limits_{q}^{\;}\left[\frac{5}{3}{\varepsilon^*}_{kq}(\rho_q,T)-{\varepsilon}_{kq}(\rho_q,T)\right]\\
   +\frac{t_0}{2}\left(1+\frac{x_0}{2}\right)\rho^2+\frac{t_3}{12}\left(1+\frac{x_3}{2}\right)(\alpha+1)\rho^{\alpha+2}\\
   -\frac{t_0}{2}\left(x_0+\frac{1}{2}\right)\sum\limits_{q}^{\;}{\rho_q}^2-\frac{t_3}{12}\left(\frac{1}{2}+x_3\right)(\alpha+1)\rho^\alpha\sum\limits_{q}^{\;}{\rho_q}^2\\
   C(\beta+1)\rho^\beta{\rho_p}^2+C_s(\eta-1)\rho^\eta,
  \end{array}
\end{equation}
with
\begin{equation}
\label{eq7}
  \begin{array}{l}
   \varepsilon_{kq}=\frac{{m^*}_q}{m}\frac{1}{\beta}\frac{2g}{\sqrt{\pi}}{\lambda_q}^{-3}F_\frac{3}{2}(\eta_q),\\
   {\varepsilon^*}_{kq}=\frac{1}{\beta}\frac{2g}{\sqrt{\pi}}{\lambda_q}^{-3}F_\frac{3}{2}(\eta_q),\\
   \eta_q(\rho_q,T)={F_\frac{1}{2}}^{-1}\left(\frac{\sqrt{\pi}}{2g}{\lambda_q}^3\rho_q\right),\\
   C\rho^\beta=\frac{4\pi}{5}e^2R^2,\\
   C_s\rho^\eta=\frac{4\pi{r_0}^2\sigma}{V^\frac{1}{3}}\rho^\frac{2}{3},
  \end{array}
\end{equation}
where $m$ and $m^*$ are the mass and effective mass respectively, $T$ and $\rho$ are the temperature and density, $q$ type of particle ($q$=proton,neutron), $F$ is the Fermi integral, $\lambda=\sqrt{\frac{2\pi\hbar^2}{m^*T}}$ is the average de Broglie thermal wavelength, $g=2$ is the spin degeneracy factor, $t_0$, $t_3$, $x_0$, $x_3$ and $\alpha$ are the Skyrme force parameters, $\beta=\frac{1}{T}$, $C\rho^\beta$ and $C_s\rho^\eta$ are the approximate Coulomb and surface effects for a finite uniform sphere of radius $R=r_0A^\frac{1}{3}$ with total charge $Z$ ($U_c=\frac{3}{5}\frac{e^2Z^2}{RV}$).
From Eqs. (\ref{eq5}) and (\ref{eq6}) for the symmetric nuclear matter with the isospin independent effective mass in the case $T=0$ at normal density $\rho_0$, one gets for $\delta$:
\begin{equation}
\label{eq8}
 \begin{array}{l}
   \delta=\frac{2}{3}\frac{1}{{\rho_0}^2}\\
   \times\frac{-33t_0-160W{\rho_0}^{-1/3}+t_3(1+\alpha){\rho_0}^\alpha\frac{1}{12}\left(7(3\alpha+6)-3(3\alpha+6)^2\right)}{\left(15t_0+\frac{1}{12}t_3(1+\alpha)\left((3\alpha+6)-(3\alpha+6)^2\right)\right)^2}\sigma_\infty,
 \end{array}
\end{equation}
where
\begin{equation}
\label{eq9}
   W=\frac{h^2}{10m}\left(\frac{3}{8{\pi}g}\right)^\frac{2}{3}\left(\frac{5-3\frac{m*}{m}}{\frac{m*}{m}}\right).
   \end{equation}

\textit{Results and Discussion.---}
To check the suggested approach, we have calculated the values of the Tolman length $\delta$ for different effective interaction, \textit{SLy6, SkM*}, and \textit{SV-min} [Tab. (\ref{tab1})] with the surface tension of the planar interface at $T=0$ adopted to be $\sigma_\infty=1.1$ Mev/fm$^2$ \cite{Ravenhall1972, *Ravenhall1983}.
\begin{table}
 \caption{Set of Skyrme parameters and correspondent nuclear properties used in this paper \cite{Chabanat1998, Klupfel2009}}
 \label{tab1}
 \begin{ruledtabular}
  \begin{tabular}{lccc}
  \textrm{Skyrme forces}&\textrm{SkM* \cite{Chabanat1998}}&\textrm{SLy6 \cite{Chabanat1998}}&\textrm{SV-min \cite{Klupfel2009}}\\
  \colrule
     $K$ (MeV) & 216.6 & 229.8 & 222.0\\
     $m*/m$ & 0.79 & 0.69 & 0.95\\
     $t_0$ (MeV$\cdot$fm$^3$)& -2645.0 & -2479.50 & -2112.248\\
     $x_0$ & 0.09 & 0.825 & 0.243886\\
     $t_3$ (MeV$\cdot$fm$^{3(1+\alpha)}$)& 15595.0 & 13673.0 & 13988.567\\
     $x_3$ & 0.0 & 1.355 & 0.258070\\
     $\alpha$ & 1/6 & 1/6 & 0.255368\\
  \end{tabular}
 \end{ruledtabular}
\end{table}
The obtained values appear to be negative for all of the chosen parametrizations [Tab. (\ref{tab2})] which means that the surface of tension is located closer to the liquid phase.
\begin{table}
 \caption{Tolman's length $\delta$ for different Skyrme force parametrizations and from \cite{Kolomietz2013} for the SkM* force }
 \label{tab2}
 \begin{ruledtabular}
  \begin{tabular}{lcccc}
  \textrm{$\;$}&\textrm{SkM*}&\textrm{Sly6}&\textrm{SV-min}&\textrm{\cite{Kolomietz2013}}\\
  \colrule
  Tolman length $\delta$ (fm) & -0.8869 & -1.5600 & -0.5512 & -0.3703  \\
  \end{tabular}
 \end{ruledtabular}
\end{table}
Those results  and in the order of magnitude with the results obtained from the Gibbs-Tolman approach \cite{Kolomietz2012, Kolomietz2013}.
All the values obtained within the present approach agree in sign with those, calculated in Refs. \cite{Kolomietz2012, Kolomietz2013} from the Gibbs-Tolman formalism applied to the charged Fermi-Liquid droplet. As for the absolute value they are slightly higher but consistent in the order of magnitude. Even though the authors of \cite{Kolomietz2012, Kolomietz2013} performed a detailed analysis of the positions of the surface of tension and equimolar surface in the nuclei in order to calculate the length, the result $\delta=-0.3703$ fm seem to underestimate the Tolman length, as it is smaller than the internucleon distance $r_0\sim0.7$ fm.

As seen from Tab. (\ref{tab2}) the obtained value for $\delta$ in the case of $SkM^*$ parametrization, introduced to account for the surface properties of nuclear matter, is close to the distance between the nucleons,

One can see from Tab. (\ref{tab2}) that the suggested approach is very sensitive to the Skyrme force parametrization and the results can differ more than by a factor of two. It seems that there can be two different reasons for this discrepancy. First is using the different phenomenological inputs when calibrating the nuclear effective energy functionals. The second is constant surface tension coefficient $\sigma_\infty$ used for the calculations when it can be different depending on the energy functional used. Although that question requires further studies it is possible provide the brief analysis of the observed difference at this step. The chosen parametrizations represent the three typical sets.

$SkM*$ is a representative of the group of Skyrme forces that were developed with an explicit study of surface energy and fission barriers \cite{Bartel1982}. Using fission barriers in $^{208}$Pb and surface coefficients $a_{surf}$ as the phenomenological input data has much improved the description of surface effects along the high-precision description of nuclear ground states with the $SkM*$. Therefore, it looks quite reasonable that the results for the Tolman length obtained for that parametrization are very close to the  internucleon distance $r_0\sim0.7$ fm, in accord with the theoretical predictions of Refs. \cite{Gorski1989, Fisher1984}.

Next, $SLy6$ is a parametrization especially designed for neutron rich matter and neutron stars \cite{Chabanat1998,Chabanat1998a}. The force aimed to be used in the astrophysical applications and it is not surprising that the Tolman length $\delta$ evaluated in this work for the symmetric matter is less consistent with the theoretical predictions for its value in comparison with the $SkM*$ force. The reason for that force to give the overestimated value may be in bigger difference in between the neutron and proton distributions in the neutron rich matter that and, therefore, the possible increase in the distance between the equimolar surface and surface of tension.
As for the third $SV-min$ force it is the recent force that was constructed using a large pool of spherical nuclei as well as some detailed observables such as neutron skin, isotope shifts, and super-heavy elements \cite{Klupfel2009}. The r.m.s. errors in the charge distribution formfactor, radius and surface thickness for that force are very close to those of the $SkM*$ force \cite{Erler2011}. That can give the explanation for the similar deviation of the Tolman length from the theoretical predictions for this two forces.
In summary it seems the suggested approach can give realistic values for the Tolman's $\delta$ correction that can be used as a test for the validity of each parametrization in treating the surface properties of the nuclei.

\begin{figure}
\includegraphics[scale=1.0]{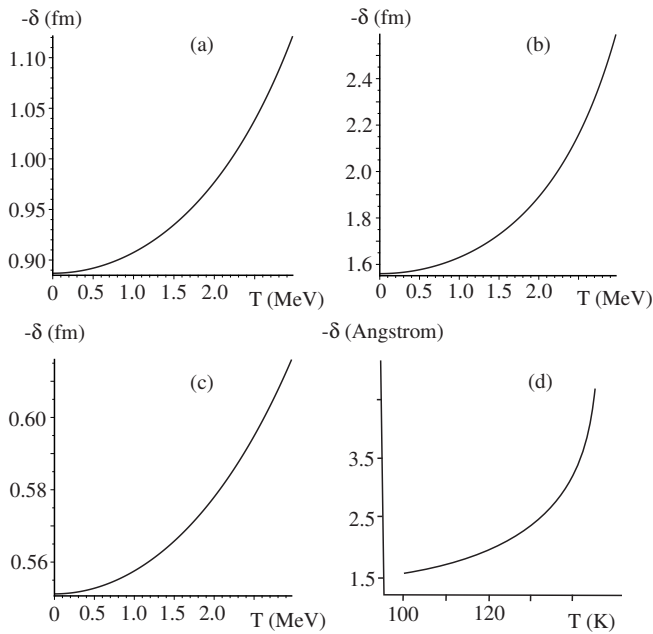}
\caption{Temperature dependence of Tolman length $\delta$. (a), (b), (c) for the nuclear matter with the parametrizations SkM*, SLy6 and SV-min respectively. Initial values $\rho_0$ correspond to the equilibrium condition $P(\rho_0,T)=0$; (d) - for the ordinary liquid $Ar$ \cite{Bulavin1998}} \label{fig2}
\end{figure}

The temperature dependence $\delta(T)$ is shown in Fig. \ref{fig2}.
One can see that the correction term in nuclear matter [Fig. \ref{fig2} (a), (b) and (c)] increases with temperature for all the studied nuclear forces parametrizations. All the curves far away from the critical point can be approximated by the equation
\begin{equation}
\label{eq10}
   \delta(T)=\delta(0)(1+aT^b),
   \end{equation}
where $a$ and $b$ are free parameters that slightly vary for all the forces. Unfortunately, we have not yet succeeded in deriving this simple approximation analytically and can not explain the physics of such a behavior. This needs some further studies. At the same that qualitative picture corresponds to the data for ordinary liquids [Fig. \ref{fig2} (d)] known in the literature \cite{Bulavin1998} and makes it especially important for the heavy ion collision experiments, where the yield of fragments is exponentially dependent on surface tension, and the nuclear matter is at high temperatures.

\textit{Conclusions.---}
In this paper we have calculated the curvature correction term in the surface tension from the nuclear equation of state for the $SLy6$, $SkM*$ and $SV-min$ parametrizations. The obtained results show the importance of that correction for light nuclei.

To summarize, our study shows that the present approach makes possible calculations of the Tolman length $\delta$ from a simple thermodynamic equations. The obtained values are consistent with the existing data and with the theoretical predictions. The temperature dependence $\delta(T)$ for the nuclear matter shows the same behaviour as that of ordinary liquids. The possibility to evaluate the temperature dependence of the curvature correction term makes the suggested approach useful in analyzing the results of heavy ion collision experiments and in calculating yields of light fragments. The present approach, based on a minimal set of assumptions, provides a simple and reliable way to calculate the curvature correction term, and it may be used to study the properties of light nuclei and of complicated nuclear processes sensitive to the surface tension.

\begin{acknowledgments}
This work was supported by the National Natural Science Foundation of China under Grants No. 11161130520 and 11025524, the National Basic Research Program of China under Grant No. 2010CB832903, and the European Commissions 7th Framework Programme (FP7-PEOPLE-2010-IRSES) under Grant Agreement Project No. 269131. L.J. was supported by the grant {\textquotedblleft}Matter under extreme conditions{\textquotedblright} of the National Academy of Sciences of Ukraine
\end{acknowledgments}

\nocite{*}
\bibliography{DeltBase}

\end{document}